\documentclass[11pt,a4paper]{article}

\usepackage[utf8]{inputenc}
\usepackage{amsmath}
\usepackage{amssymb}
\usepackage{type1cm}
\usepackage{graphicx}
\usepackage{multirow}
\usepackage{braket}
\usepackage{titlesec}

\newcommand{\abs}[1]{ \left| {#1} \right| }

\renewcommand{\set}[1]{\left\lbrace {#1} \right\rbrace}
\newcommand{\lp}{\left(}
\newcommand{\rp}{\right)}
\newcommand{\pr}{^{\prime}}

\DeclareMathOperator\range{range}

\newcommand{\ifaf}{if and only if }

\newcommand{\ea}{\textit{et al. }}
\newcommand{\etc}{\textit{etc. }}
\newcommand{\eg}{\textit{e.g. }}

\newcommand{\nmm}{n_{\max}-n_{\min}}

\title{The Effect of `Probability Skew' in Bell-test Experiments}
\author{Dale Hodgson}
\date{}

\titleformat*{\section}{\large\bfseries}

\begin{document}

\maketitle
\begin{abstract}
We consider typical experiments that use Bell-inequalities to test local-realist theories of quantum mechanics and gain insight into how certain results can be obtained. We see that results against local-realism arise from some `quantum skew' of the correlation between entangled qubit pairs. Furthermore, we find some conditions necessary for a conclusion against local-realism. Finally we show that the problem of `no-signalling' that arises in these experiments cannot be reduced to arbitrary experimental accuracy.
\end{abstract}

\section*{Introduction}
The testing of Bell-inequalities is the favoured method of investigating the relationship between local-realism and quantum mechanics. Significant results have been achieved in this area, but further work is still ongoing.

Realism is the assumption that the physical universe \emph{exists} (has some `element of reality') independent of any measurement or observation. Locality is the assumption that physical influences cannot travel faster than the speed of light --- the cornerstone of special relativity. In their 1935 paper \cite{epr}, Einstein, Podolsky and Rosen noted the disparity arising between the assumptions of local-realism and the behaviour of entangled quantum objects. John Bell's later work on that idea \cite{bell} lead to experimentally testable results, now known as Bell-inequalities, that can test whether certain quantum mechanical systems obey the ideas of local-realism.

A popular type of Bell-test experiment involves two entangled qubits being measured under space-like separation sufficient to exclude relativistic interaction and noting the correlation of the two sets of results. The correlations become a test value which, under the assumption of local-realism, has a theoretical bound. Hence such experiments can verify or contradict any local-realist theory for the behaviour of quantum objects.

The testable theoretical bound arises from consideration of a `correlation coefficient'. If $a$ is a measurement examining the state of the first qubit, and likewise $b$ for the second, then the correlation coefficient for the joint outcome $ab$ is
\begin{align*}
E_{ab} & = probability~of~correlation - probability~of~anti\text{-}correlation \\
& \text{realised experimentally as: } \frac{n(corr)_{ab} - n(anti\text{-}corr)_{ab}}{n(ab~trials)}\text{.}\\
\end{align*}

Bell's work shows that if some `hidden variable' that could not be directly measured exists (the assumption of \emph{reality}), then the inequality:
\begin{align}
E_{ac} - E_{ba} - E_{bc} \leqslant 1 \label{bell1}
\end{align}
should hold (in a set-up with perfect anti-correlation).

The later work of Clauser, Horne, Shimony and Holt \cite{chsh} generalised this to allow for non-perfect anti-correlation in what is now known as the CHSH inequality. With two measurements prepared for each qubit, chosen such that measurement $1$ is precisely the complement of measurement $0$ (\eg measures of spin in perpendicular directions).
\begin{align}
E_{00}+E_{01}+E_{10}-E_{11} \leqslant 2 \label{chsh}
\end{align}

Recent work by Hensen \textit{et al.} \cite{hensen1} \cite{hensen2} have shown statistically significant violation of the CHSH-inequality, suggesting to us that there is no local-realist theory that completely describes quantum mechanics.

\section*{A Bell-test experiment using CHSH}
A general Bell-test experiment produces results with \emph{eight} independent variables: $N$ the total number of trials, $a, b, c$ the number of trials with setting configurations $00, 01, 10$, respectively, and the number of correlated results under each setting ($n_{00} , n_{01} , n_{10}, n_{11}$). (See Table 1.)

\begin{table}[h]
\caption{General results from a CHSH experiment}
\centering
\resizebox{\columnwidth}{!}{
\begin{tabular}{c|c|c|c|c}
Setting & 00 & 01 & 10 & 11 \\ 
\hline
\rule[-1ex]{0pt}{4ex} No. of Trials & $a$ & $b$ & $c$ & $d=N-a-b-c$ \\
\rule[-1ex]{0pt}{4ex} No. of Correlated results & $n_{00}$ & $n_{01}$ & $n_{10}$ & $n_{11}$ \\ 
\rule[-1ex]{0pt}{4ex} No. of Anti-correlated results & $a-n_{00}$ & $b-n_{01}$ & $c-n_{10}$ & $d-n_{11}$ \\ 
\rule[-1ex]{0pt}{4ex} $p(corr)$ & $\frac{n_{00}}{a}$ & $\frac{n_{01}}{b}$ & $\frac{n_{10}}{c}$ & $\frac{n_{11}}{d}$ \\ 
\rule[-1ex]{0pt}{4ex} $p(anti\text{-}corr)$ & $1-\frac{n_{00}}{a}$ & $1-\frac{n_{01}}{b}$ & $1-\frac{n_{10}}{c}$ & $1-\frac{n_{11}}{d}$ \\ 
\rule[-1ex]{0pt}{4ex} $E=p(corr)-p(anti\text{-}corr)$ & $\frac{2 n_{00}}{a}-1$ & $\frac{2 n_{01}}{b}-1$ & $\frac{2 n_{10}}{c}-1$ & $\frac{2 n_{11}}{d}-1$ \\ 
\end{tabular}
}

\end{table}

Giving test value: 
\[
S= E_{00} + E_{01}+E_{10}-E_{11} = 2 \left( \frac{n_{00}}{a} + \frac{n_{01}}{b} + \frac{n_{10}}{c} - \frac{n_{11}}{d} - 1 \right).
\]

Local-realism predicts $S \leqslant 2$, so an experiment disproving local-realist explanations of quantum entanglement will violate the CHSH-inequality --- showing $S > 2$.

\section*{The Effect of `Quantum probability skew'}

Suppose that the probabilities of correlation under each experimental setting are equal (to $P$ say). Then $S=2 \lp 2P -1 \rp$; in which case $S>2 \Leftrightarrow P > 1$. Hence uniformity of these correlations will \emph{never} violate the CHSH-inequality --- the nature of quantum mechanics lends some `unnatural skew' to these results.

Similarly, we may examine Bell's original inequality \eqref{bell1}. Let $n_{ac}$ denote the number of correlated results under setting $ac$, $N_{ac}$ the total number of trials under setting $ac$, $n_{ba}$ the number of correlated results under measurement setting $ba$ \etc Then correlation values take the form $E_{ac}=\frac{n_{ac}-\lp N_{ac}-n_{ac} \rp}{N_{ac}}=\frac{2 n_{ac}}{N_{ac}}-1$ and Bell's inequality asserts that a local-realist theory implies the following inequality:
\[
\frac{n_{ac}}{N_{ac}} - \frac{n_{ba}}{N_{ba}} - \frac{n_{bc}}{N_{bc}} \leqslant 0 \text{ .}
\]
Again, if these three fractions (each equal to the probability of correlation under a given setting) are all equal, then the inequality will always be satisfied.

\section*{Approximating this Experiment for Large Number of Trials}

Since the experimental settings are determined randomly, we would expect that the number of trials under each setting would tend to a uniform distribution for suitable large $N$. Supposing this is the case, let $a=b=c=\frac{N}{4}$, then we get 
\[
S=\frac{8}{N}\lp n_a + n_b + n_c - n_d - \frac{N}{4} \rp \text{ .}
\]

Let us denote the range of correlation counts $\sigma$.
\[
\sigma := \range\set{n_a,n_b,n_c,n_d}= \nmm
\]

Where $n_{\max} = \max\set{n_a,n_b,n_c,n_d}$ and $ n_{\min} = \min \set{n_a,n_b,n_c,n_d}$. Let us only consider $\sigma \geqslant 1$ (for if $\nmm = 0$, then we will always get $S<2$). Furthermore, we must have that $n_a+n_b+n_c+n_d \leqslant N$; consequently, it is clear that we must have $n_{\min}\leqslant \frac{N}{4}$.

Let us denote $(n_a+n_b+n_c-n_d)$ by $S \pr$. Then we will observe violation of the CHSH inequality ($S>2$) \ifaf $S \pr > \frac{N}{2}$. Note also that $S \pr$ is bounded above and below by $S \pr _{\max} = 3 n_{\max}-n_{\min}=2n_{\min}+3\sigma$ and $S \pr _{\min}=3n_{\min}-n_{\max}=2n_{\min}-\sigma$.

Violation of the CHSH inequality is only possible if $S \pr _{\max} > \frac{N}{2}$, that is to say: $2n_{\min}+3\sigma>\frac{N}{2}$ is \emph{necessary} (but not sufficient) for violation. If we desire violation of a certain magnitude $\delta$, then we may say that $S \pr _{\max} > \frac{N}{2}+\delta \Leftrightarrow 2n_{\min}+3\sigma>\frac{N}{2}+\delta$ is necessary. From this we may see that for a violation of magnitude $\delta$, the `quantum skew' (range of correlation counts) $\sigma$ must be strictly greater than $\frac{\delta}{3}$:

\[
\begin{array}{rcc}
\sigma >\frac{N}{6}+\frac{\delta}{3}-\frac{2}{3}n_{\min} & \multirow{2}{*}{$\bigg \rbrace$ } & \multirow{2}{*}{$\Rightarrow \sigma > \dfrac{\delta}{3} \text{ .}$}\\
 n_{\min} \leqslant \frac{N}{4} & ~ & ~
\end{array}
\]

If we desire $S>2+ \Delta$, then we are in fact asking that $S \pr > \frac{N}{2}+\frac{N \Delta}{8}$, so $\delta = \frac{N \Delta}{8}$; hence $\sigma > \frac{N \Delta}{24}$ is necessary. 

\section*{The No-signalling Problem}

A crucial factor in these experiments is that the two measurements must be made \emph{independent} of one another. Khrennikov's analysis of the first Hensen \ea  experiments \cite{khr} noted that the independence of the measurements could be verified by analysing marginal probabilities available from the data.  Bednorz \cite{bednorz} also reviewed recent experiments in a similar way.

That is to say, if the measurements are truly independent, then we should observe:
\begin{align*}
p(corr|s_1=0) & = p(corr | s_1 = 0 \wedge s_2=0)\\
& = p(corr | s_1 = 0 \wedge s_2 = 1) \text{ ,}
\end{align*}
and similarly for other settings.

In the experimental set-up we are considering, all such probability differences look like
\[
p(corr|0s_b)-p(corr|00)=\frac{n_a \! + \! n_b}{a \!+ \! b}-\frac{n_a}{a}=\frac{an_b \! - \! bn_a}{a(a \! + \! b)}.
\]

The family of all these equations can be described by $\frac{\alpha n_{\beta}-\beta n_{\alpha}}{ \alpha (\alpha + \beta) }$ and $\frac{\beta n_{\alpha} - \alpha n_{\beta}}{ \beta (\alpha + \beta) }$ where $\alpha , \beta \in \set{a,b,c,d}$ and $\alpha \neq \beta$.

We can say that we have achieved experimental results with a good no-signalling accuracy if all these probability differences are small. All such probabilities will be `small' (to some $\varepsilon$), if
\[
\frac{\abs{\alpha n_{\beta} - \beta n_{\alpha}}}{\alpha + \beta} < \varepsilon \cdot \min \set{\alpha , \beta}.
\]

In the case of uniform experimental test settings considered above ($a=b=c=\frac{N}{4}$), this condition becomes: $\abs{n_{\alpha}-n_{\beta}} < \frac{ \varepsilon N}{2} ~ \forall \alpha , \beta$. This is satisfied if 
\[
\nmm < \frac{\varepsilon N}{2} \text{ .}
\]

So $\sigma < \frac{\varepsilon N}{2}$ is necessary to achieve results with a good no-signalling accuracy. Since $\sigma \geqslant 1$ is necessary for a CHSH-inequality violation, $N > \frac{2}{\varepsilon}$ trials must be performed if an experiment violating the CHSH-inequality is to achieve no-signalling tolerance of accuracy $\varepsilon$.

Combining this with our knowledge that $\sigma>\frac{N \Delta}{24}$ is necessary for violation of the CHSH inequality to magnitude $\Delta$, we conclude that 
\[
\varepsilon > \frac{\Delta}{12}
\]
is a limit on our no-signalling tolerance --- independent of the number of trials we perform.

\section*{Acknowledgements}
With thanks to my PhD supervisor Dr Vladimir Kisil for putting me on to this topic and his advice while researching and writing.

\bibliographystyle{hplain}
\bibliography{probability_skew_in_bell_experiments}

\end{document}